\documentclass[aps,12pt,showpacs]{revtex4}

\usepackage[english]{babel}
\usepackage{amsmath,amssymb,amsbsy,amstext}
\usepackage[dvips]{graphicx}

\newcommand{\be}{\begin{eqnarray}}
\newcommand{\ee}{\end{eqnarray}}
\newcommand{\bdm}{\begin{displaymath}}
\newcommand{\edm}{\end{displaymath}}
\newcommand{\ds}{\displaystyle}   
\newcommand{\ba}{\begin{array}}
\newcommand{\ea}{\end{array}}
\newcommand{\pa}[1]{\left(#1\right)}
\newcommand{\paq}[1]{\left[#1\right]}

\newcommand{\az}{\mathcal{S}}
\newcommand{\lag}{\mathcal{L}}

\newcommand{\mg}{\mathcal{G}}
\newcommand{\dpa}{\partial}
\newcommand{\al}{\alpha'}
\newcommand{\si}{\sigma}
\newcommand{\erf}{{\rm erf}}
\newcommand{\ket}[1]{\left|#1\right\rangle}
\newcommand{\bra}[1]{\left\langle#1\right|}

\begin{document}

\begin{flushright}
\normalsize{hep-th/0606275}
\end{flushright}

\title{Graviton production from D-string recombination and annihilation}
\author{Jean Louis Cornou\protect\( ^{(1)}\protect\), 
Enrico Pajer\protect\( ^{(2)}\protect\) and 
Riccardo Sturani\protect\( ^{(3)}\protect\)}
\affiliation{
(1) \'Ecole Normale Superieure de Paris, France,\\
(2) Ludwig Maximilians Universit\"at M\"unchen, Germany,\\
(3) Department of Physics, University of Geneva, Switzerland and INFN, Italy\\
\texttt{e-mail: cornou@clipper.ens.fr, enrico.pajer@physik.lmu.de, 
riccardo.sturani@physics.unige.ch}}

\begin{abstract}
Fundamental superstrings (F-strings) and D-strings may be produced at high 
temperature in the early Universe. Assuming that, we investigate if any of the 
instabilities present in systems of strings and branes can give rise to a 
phenomenologically interesting production of gravitons. 
We focus on D-strings and find that D-string recombination is a far too weak 
process for both astrophysical and cosmological sources.
On the other hand if D-strings annihilate they mostly produce massive closed
string remnants and a characteristic spectrum of gravitational modes is
produced by the remnant decay, which may be phenomenologically interesting
in the case these gravitational modes are massive and stable.
\end{abstract}


\maketitle

\section{Introduction}

After the dismissal of cosmic string as candidate seeds for the formation 
of structure, they received a fair share of attention in the last few years 
from both the theoretical and observational perspective.\\
Scenarios of inflation involving branes have become popular 
recently \cite{Kachru:2003sx,Jones:2002cv}, because of the appealing feature
of providing a working model of 
inflation based on a fundamental theory on one side, and because they can
represent a phenomenological handle for string theory on the other.
In this kind of scenarios the production of D-strings as cosmic defects, 
by Kibble mechanism, is a generic prediction \cite{Sarangi:2002yt}, even if 
their (meta-)stability is not at all ensured \cite{Copeland:2003bj}.

In general, cosmic defects like monopoles, strings and walls can be problematic
as their red-shift is slower than the one of radiation: if present in the
early Universe and if cosmologically stable they tend to overclose the 
Universe, unless they are diluted by some release of entropy into the 
Universe after their creation. While this is certainly true for stable 
monopoles, whose annihilation rate is negligible for standard choice of 
parameters \cite{Preskill:1979zi}, in the standard scenario cosmic strings 
loose energy by chopping off small loops and finally reach \emph{scaling} 
solutions on which they make up a constant fraction of the energy density of 
the Universe during both the era of radiation and matter domination 
\cite{Vilenkin:book}. This allows the density of long strings to be always 
roughly few per Hubble volume, while the small loops disappears by emitting 
gravitational radiation.\\
The situation is less clear for domain walls, as the existence of scaling
solutions and the time required for the solution to fall into these kind of
attractors in presently under investigation 
\cite{Avelino:2005peAvelino:2006ia}.

In this work we concentrate on cosmic strings. In particular we want to 
identify cosmic strings with the D-strings of string theory.

The production of lower dimensional defect can result for instance from D$p$ 
${\overline{\rm D}}p$-brane annihilation, which will produce D$(p-2n)$-branes 
as remnants. 
In a type II theory D$p$-branes are BPS only if $p$ has a definite parity 
(even in IIA, odd in IIB), D$p$-branes with the ``wrong'' dimensionality 
will decay immediately (on time scale given by the string scale).

Actually in type I string theory non BPS D0-branes can be stable,
because the unstable mode is projected out by the
orientation projection taking type IIB into type I. Still it has to be remember
that the presence of a type I excitation spectrum is related to the presence of
orientifolds. Even if the starting setup is endowed with a space filling
orientifold, since some of the dimensions may get small during
cosmological evolution, T-dualities over those directions getting smaller
than the string length confine the orientifold hyperplane to subspace of 
strictly positive codimension, so the bulk physics, away from the
orientifolds, will be again unoriented, i.e. of type IIA or IIB. Thus even 
taking a full type I configuration as a starting point, the general 
assumption that only the BPS D$p$-branes are stable is justified on general 
grounds.

Strictly speaking BPS D-strings are present in type IIB and type I
theories and not in type IIA, but we can consider D$p$-branes with
arbitrary $p$ wrapped around $(p-1)$ dimensional cycles to be D-strings from 
the 4 dimensional point of view, allowing a variety of values for the tension 
of the effective D-string. 
For instance, the tension of a D$p$-brane is $\mu_p\propto
\al^{-(p+1)/2}g_s^{-1}$, (being $\al\equiv l_s^2$ the square of the string
fundamental length) 
and assuming that a D$(2+n)$-brane wraps around a $n$ dimensional flat
torus with radii $R_1,\ldots ,R_n$ the effective tension $\mu$ of the resulting
4 dimensional D-string is 
\bdm
\mu=\frac 1{2\pi\,{l_s}^2g_s}\prod_{i=1}^n\frac{R_i}{l_s}\,.
\edm
Moreover the internal dimensions do not need to be flat, but can be warped (\`a
la Randall-Sundrum \cite{Gogberashvili:1998vxRandall:1999vf}, for instance), so
that their tension can be redshifted (or blue-shifted) by huge factors and in 
principle it can take any value.

One important and not completely settled issue is the stability one.
The (effective) D-strings are generically created but for them to play a role
in cosmology they must be stable, or at least metastable over cosmological
time scale. 

An instability already pointed out in \cite{Witten:1985fp} is the following.
D-strings couple electrically to a 2-form, whose dual in 4 dimensions is a
scalar, axion-like field. Once supersymmetry is broken the axion acquires a 
periodic potential with several degenerate minima, so the axion is expected to 
condense to different minima in different regions of space, separated by domain
walls. Finite domain walls are bound by D-strings, and the domain wall
tension, unless it is exceedingly small, will make D-strings shrink and 
collapse.

Another source of instability is breakage, which is present whenever systems
of D$p$ and D$q$-branes are present at the same time. Let us define $\nu$ as 
the 
number of dimensions in which the D$p$ branes are extended and the D$q$ are
not or vice-versa, i.e. the total number of Neumann-Dirichlet (ND) plus DN 
dimensions. If $\nu=2$ the 
axion symmetry is Higgsed, there are no axion domain walls, so the branes can 
break. Strings can in general break or annihilate if they meet
a brane with $\nu=2$, if $\nu=4$ or $\nu=8$ the system is stable and 
supersymmetric, but we fall back into the domain wall problem when 
supersymmetry is broken (if $\nu=6$ branes repel each others).

Since the branes come with orientation, meeting a brane with
opposite orientation they can also annihilate, 
whereas if they have exactly the
same orientation and $\nu=0,4,8$ the full system is still BPS. 
Of course intermediate situation are possible, when branes meet at angle. 
A description of the microscopic mechanism leading to recombination is given
in \cite{Hashimoto:2003xz,Hanany:2005bc}.

The stability issue is thus far from settled but it can be solved
in specific models. Here we will focus on the effects triggered by
two of the aforementioned sources of instability, the meeting of a D-string
at angles and the annihilation of D-strings.
This kind of processes can be addressed in the context of string theory and the
dynamic of the instability can be studied through the time evolution of a 
specific tachyonic mode of the system \cite{Sen:2004nf}.\\
Since everything is coupled to gravity such instabilities can be expected to 
give rise to some gravitational radiation, which we study quantitatively in
the present work.
Modeling the string sources with an effective energy momentum tensor
the amount of emitted gravitational radiation can be computed, but a
negligible result is found.

Actually, rather than the emission by a time dependent dynamics we can consider
the closed string remnant left by a brane-anti brane annihilation
\cite{Karczmarek:2003xm}, whose decay can be an 
interesting source of gravitational radiation, as we will show.

It is also worth noticing that no cosmic string has ever been observed
directly and that the cosmic microwave background constrains cosmic
string tension $\mu$ to be $G_N\mu< 10^{-7}$, where $G_N$ is the Newton 
gravitational constant.

The outline of the paper is the following.
In sect.~\ref{recmb} the salient features of the analysis of
\cite{Hashimoto:2003xz} of the D string reconnection are reproposed and the GW 
emission in such a process is computed, then generalised to a cosmological 
setup like in \cite{Damour:2001bk}.
In sect.~\ref{annhi} the annihilation process between a D string and a
$\overline{\rm D}$ string is analysed, finding again a negligible
gravitational radiation emission. 
In sect.~\ref{remnt} we study the gravitational radiation emitted by the
decay of the massive closed strings which constitute the remnant of the 
D $\overline{\rm D}$-brane annihilation and conclusions are finally drawn in 
sec.~\ref{concl}.

\section{Gravitational radiation from string-string reconnection} 

\label{recmb}
The reconnection process of D-strings meeting at an angle and zero relative 
velocity has been studied in \cite{Hashimoto:2003xz}. Adding a relative 
velocity between the strings boils down to introducing a non one (and non zero)
probability of interaction \cite{Jackson:2004zg,Hanany:2005bc}.

Stacks of parallel D-branes admit a low energy description in terms of a 
$U(N)$ super Yang Mills theory, where $N$ is the number of D-branes into play.
The bosonic excitations of the F-strings stretched between the D-strings are 
gauge vectors (scalars) from the point of view of the D-string Poincar\'e 
group, if polarised along (orthogonally to) the brane. The system is described 
by a 1+1 dimensional Yang Mills theory coupled to as many scalars as transverse
direction. The scalar fields parametrise the displacements of the brane in the 
orthogonal directions.

\subsection{Almost parallel D-strings}

\subsubsection{Reconnection dynamics}

Following the analysis of \cite{Hashimoto:2003xz}, let us consider a 
system made of 2 almost parallel D-strings at an angle $\gamma$.
The low energy dynamics of the system is described by the action
\be \label{action}
\az=-\frac\mu 2\pa{2\pi\al}^2{\rm Tr}\int dxdt\pa{\frac 14 F_{\mu\nu}F^{\mu\nu}
+\frac 12D_\mu\Phi_iD^\mu\Phi^i}\,,
\ee
where $\mu=0,1$ and $i\in \{2\ldots d\}$ runs over the dimension orthogonal to 
$t$ and $x$.
\begin{figure}
  \centering
  \includegraphics[width=.8\linewidth]{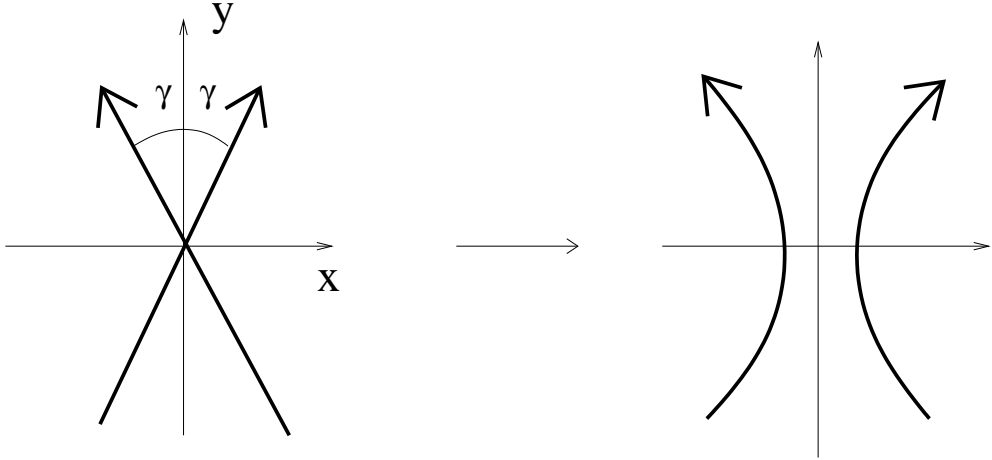}
  \caption{Two D-branes at an angle $\gamma$ before and after recombination.}
\end{figure}

In order to describe two D-strings at an angle $\gamma$ the previous
action (\ref{action}) has to be expanded around the background $\Phi^{(b)}_i$ 
$A_\mu^{(b)}$
\be \label{back}
\Phi_i^{(b)}=\delta_{i,2}\ q x\,\si^3\,,\qquad A_\mu^{(b)}=0\,,
\ee
with $q\equiv\tan\gamma/(\pi\al)$ and $\si^i$ the Pauli matrices which form the
algebra of $SU(2)$, plus the fluctuations $\Phi^{(f)}$, $A_\mu^{(f)}$
\be \label{fluct}
\Phi_i^{(f)}=\delta_{i,2}\pa{\xi(t,x)\,\si^3+\phi(t,x)\,\si^1}\,,\qquad 
A_\mu^{(f)}=\delta_{\mu,1}\ a(t,x)\,\si^2\,.
\ee
To lowest non trivial order in interactions, the resulting Lagrangian density 
is, \footnote{The set of fluctuations 
$\phi_2\propto\si^2$, $A_1\propto\si^1$ decouple from $\phi_2\propto\si^1$, 
$A_2\propto\si^2$ and its Lagrangian is another copy of (\ref{lagdd}).}
\be \label{lagdd}
\ba{rl}
\ds\lag\simeq&\ds\frac 12 \dot a^2+\frac 12\pa{\dot\phi^2+\dot\xi^2}-\frac 12
\pa{{\phi'}^2+{\xi'}^2}-\frac{q^2}2-q\xi'-a\phi'\pa{qx+\xi}+a\phi\pa{q+\xi'}\\
&\ds -\frac 12 a^2\pa{q^2x^2+2qx\xi}\,,
\ea
\ee
and it takes to the equations of motion 
\be \label{aphixi}
\ba{rcl}
\ddot a +qx\phi'-q\phi+aq^2x^2&=&0\,,\\
\ddot\phi-\phi^{''}-a'qx-2aq&=&0\,,\\
\ddot\xi-\xi^{''}+a'\phi+2a\phi'+a^2qx&=&0\,,
\ea
\ee
where it is understood that a dot stands for a time derivative and a prime for
a derivative with respect to $x$.\\
We note that consistency of the equations requires that $a,\phi$ are 
of the same order, while $\xi$ is $O(a^2)$, so that the first
two equations of (\ref{aphixi}) can be solved for $a$ and $\phi$ and then the 
result plugged into the equation for $\xi$.
This justifies a posteriori the choice made in (\ref{lagdd}) to drop $O(a^3)$ 
terms as well as $O(a \xi^2 )$.

For the coupled $a,\phi$ system it is convenient to take the ans\"atze
\be
\ba{c}
a(t,x)=\sum_{n>0} A_ne^{im_n t}a_n(x)\,,\\
\phi(t,x)=\sum_{n>0}F_ne^{im_nt}a_n(x)\,,
\ea
\ee
where the equation of motions force the condition $m_n^2=(2n-1)q$.\\
Both $a$ and $\phi$ have an unstable lowest lying mode, with a Gaussian 
profile, so that neglecting the positive mass squared mode the solution can be 
written as \cite{Hashimoto:2003xz}
\be \label{solbb}
a(t,x)=A_0e^{\sqrt qt}e^{-qx^2/2}=\phi(t,x)\,.
\ee
The expression for $\xi$ is slightly more complicated, but again its non 
trivial part is concentrated in an area of size $q^{-1/2}$ around $x=0$
\be \label{xi}
\xi(t,x)=A_0^2e^{2\sqrt qt}\frac e2\sqrt\frac{\pi}q\paq{
e^{2\sqrt qx}\erf(\sqrt qx+1)+e^{-2\sqrt qx}\erf(\sqrt qx-1)
-2\sinh(2\sqrt qx)}\,.
\ee
In the previous (\ref{xi}) the integration constant of the solution to the 
homogenous part of the $\xi$ equation is fixed by requiring that the 
$\lim_{x\pm\infty}\xi=0$ independently of $t$.

To give a geometrical interpretation we remind that the $\Phi_i$'s keep trace 
of the displacements of the D-strings in the orthogonal directions, their 
eigenvalues are proportional to the locations of the D-branes in the $i$-th 
direction. In our case the distance $d_y$ along the $y$ direction between the 
2 D-strings is $d_y=2\times 2\pi\al\Phi_y$ where $\Phi_y$ is given by
\be \label{position}
\ds
\Phi_y=\frac 12
\pa{\ba{cc}
qx+\xi & \phi_0(x)\\
\phi_0(x) & -qx-\xi
\ea}\to 
\frac 12\pa{\ba{cc}
\sqrt{q^2x^2+2qx\xi+\phi_0^2} & 0\\
0 & \sqrt{q^2x^2+2qx\xi+\phi_0^2}
\ea}\,,
\ee
where in the previous (\ref{position}) a gauge transformation has been 
performed to diagonalise the $\phi$ field and only the first order in $\xi$
has been kept. The dynamics of the Yang Mills system 
tells the D-strings to recombine smoothly around the intersection point.
This analysis neglects in the Lagrangian terms of the order $a^4$ and 
higher, thus it is valid for 
\be \label{val1}
A_0e^{\sqrt qt}e^{-qx^2/2}<\sqrt q\,.
\ee 
By a shift in time we can set $A_0=\sqrt{q}$ (as we will do in the rest of the 
paper), so the solution (\ref{solbb}) can be expected not to be valid for 
$t>0$.\\
Here we neglected the effect of a relative velocity $v$ between the D-strings, 
which affects the probability of the reconnection taking place. It has been 
studied in \cite{Hanany:2005bc}, with the result that the probability $P\sim 1$
if $v\ll 1$, which takes to the condition 
\be \label{velprob}
\sqrt q e^{\sqrt qt}\ll \sqrt{q}\,.
\ee
Again this restriction suggests that this picture is a good description only 
for negative times.

\subsubsection{Energy momentum tensor}

\label{emt1}
Having understood the dynamics of the recombination via the effective action 
of the lightest string modes, it is then possible to compute the energy 
momentum tensor of the system to estimate the amount of 
gravitational radiation emitted during the process. We take the definition
\be \label{emtdef}
T^{\mu\nu}=-\frac 2{\sqrt{-g}}\frac{\delta\az}{\delta g_{\mu\nu}}\,.
\ee
Explicitly on the solutions (\ref{solbb}) the $t,x$ components of the energy 
momentum tensor are
\be \label{emt}
\ba{cl}
T_{\mu\nu}=\ds\mu 2\pi^2\al^2&\ds\pa{
\ba{cc}
-q^2/2-q\xi' & -q\dot \xi\\
-q\dot\xi & a^2q -q\xi' -q^2/2
\ea}\\
&\ds\times\paq{\delta\pa{y-qx\pi\al}+\delta\pa{y+qx\pi\al}}\delta(z)\,,
\ea
\ee
where it is understood that the strings are immersed in 4-dimensional 
space time.\\
In eq.(\ref{emt}) we recognise the constant contribution from the 
background configuration of the tilted D-strings, which is T-dual to a 
constant magnetic field on a D2-brane. The time dependent part is the source
of the gravitational radiation.
This energy momentum tensor is conserved, but it would 
\emph{not} have been conserved without the inclusion of $\xi$ in the ans\"atze
(\ref{fluct}), like in \cite{Hashimoto:2003xz}, where the solution considered 
have $\xi=0$. This should not come as a surprise as forcing the 
solution (\ref{back}) with $\xi=0$ would imply an \emph{explicit} 
breaking of Lorentz invariance, which would necessarily let no right to 
expect a conserved energy-momentum tensor. The inclusion of $\xi$ turns the 
Lorentz symmetry breaking of the background into a spontaneous one, thus 
allowing the conservation of the energy momentum tensor.

By linearising general relativity over a flat background
\be
g_{\mu\nu}=\eta_{\mu\nu}+h_{\mu\nu}\,,
\ee
we have
\be \label{wave}
\Box\bar h_{\mu\nu}=16\pi G_NT_{\mu\nu}\,,
\ee
where $\bar h_{\mu\nu}\equiv h_{\mu\nu}-1/2\eta_{\mu\nu}h^\rho_\rho$ has been 
defined and the condition $\dpa_\mu\bar h^\mu_\nu=0$ has been imposed, 
consistently with the conservation of the energy momentum tensor.
Following \cite{Weinberg:book} the energy momentum tensor can be Fourier 
transformed both in space and time 
\be
\tilde T_{\mu\nu}(\omega,\vec k)=\int T_{\mu\nu}(t,\vec x)
e^{i(\omega t-\vec k\cdot \vec x)}dt d^3x\,,
\ee
to obtain the formal solution to (\ref{wave}) as
\be
\ba{rl}
\ds \bar h_{\mu\nu}(t,x)=&\ds 4G_N\int d^3x'\frac 1{|\vec x-\vec x'|}\int 
\frac{d\omega}{2\pi}\frac{d^3k}{\pa{2\pi}^3}\tilde T_{\mu\nu}(\omega,\vec k)
e^{-i(\omega(t-|\vec x-\vec x'|)-\vec k\cdot\vec x')}\\
\simeq &\ds\frac{2G_N}{\pi x}\int\tilde T_{\mu\nu}(\omega,\omega\hat n)
e^{i\omega(t-x)}d\omega\,,
\ea
\ee
where $\hat n$ is the unit vector pointing into the direction of $\vec x$ and 
the local wave zone approximation 
$|\vec x-\vec x'|\simeq x-\vec x'\cdot\hat n$ 
has been used, with $x\equiv |\vec x|$.\\
Given (\ref{emt}) and the explicit solution (\ref{solbb})
 \be
 a(t,x)=\sqrt{q} e^{\sqrt{q}t} e^{-qx^2/2}\quad\mbox{for}\ t<0
 \ee
 one can compute 
$\bar h_{11}$, say, to be
\be \label{radiation}
\ba{rl}
\bar h_{11}(t,\vec x)=&\ds \frac 4x G_N\mu\,q^{3/2}\al^2
\pi^{3/2}
\int d\omega\frac 1{2\sqrt q+i\omega}e^{-i\omega(t-x)}
\paq{\ds e^{-\frac{\omega^2}{4q}\delta_+}+
e^{-\frac{\omega^2}{4q}\delta_-}}=\\
&\ds\frac 4xG_N\mu\,q^{3/2}\al^2\pi^{5/2}e^{2\sqrt q(t-x)}
f(t-x,\theta,\phi)\,,
\ea
\ee  
valid for $t<x$, where $f(t-x,\theta,\phi)$ takes account of the 
angular dependence of the radiation and its explicit form is
\be
f(u,\theta,\phi)\equiv
\ds e^{\delta_+^2}\paq{1+\erf\pa{\frac{\sqrt q}{\delta_+}\,u+\delta_+}}+
e^{\delta_-^2}\paq{1+\erf\pa{\frac{\sqrt q}{\delta_-}\,u+\delta_-}}\,,
\ee
where $\delta_\pm=|\sin\theta(\cos\phi\pm q\pi\al\sin\phi)|$ and $\theta,\phi$
are the usual polar angles.
Actually the details of the angular distribution are of little importance in 
view of the intensity of the radiation, as we will now show.

To compare with experimental sensitivity it is convenient to express the 
previous result in terms of the (double sided) spectral density $S_h(f)$ 
which for a short burst of (Fourier transformed) amplitude $\tilde h(f)$ 
and duration $\Delta t$ reduces to 
\be \label{shdt}
S_h(f)\simeq \tilde h^2(f)/\Delta t\,.
\ee
From the computation (\ref{radiation}) we can estimate a spectral density
function, neglecting the angular dependence in the exponential,
\be \label{spectralh}
\ba{rl}
\ds S_h(f)&\!\!\ds\simeq 10^3\frac{\pa{G_N\mu}^2q^{5/2}\al^4}{x^2}
\frac{e^{-\frac{\omega^2}{4q}}}{1+\omega^2/(4q)}\\ 
&\!\!\ds\simeq 10^{-120}{\rm Hz}^{-1}\pa{\tan\gamma}^{5/2}
\frac{e^{-\frac{\omega^2}{4q}}}{1+\omega^2/(4q)}
\pa{\frac{G_N\mu}{10^{-7}}}^2\pa{\frac{\al}{{\rm TeV}^{-2}}}^{3/2}\!\!
\pa{\frac x{1{\rm Mpc}}}^{-2}\!\!,
\ea
\ee
where redshift factors has been neglected.
The typical signal frequency is $\simeq \sqrt q/(\pi(1+z))$.
For presently working experiments $S_h\gtrsim 10^{-42}$Hz$^{-1}$ 
\cite{Abbott:2005ez}, so the result (\ref{spectralh}) is ridiculously small.
This is due to the inherently microscopic origin 
of the dynamics, whereas to generate a consistent amount of radiation a 
coherent motion of a macroscopic mass is needed. 
Let us now investigate if a cosmological population of such recombining 
D-strings may help in enhancing the signal.

\subsubsection{Cosmological rate}

\label{reccr}
One can expect that several processes like the one described in the previous
section take place in the Universe and that the signal resulting from the sum 
of all of them  is highly enhanced with respect to the one found in 
(\ref{spectralh}).

We assume the Universe evolution is the standard one after cosmic string
production, i.e. radiation domination followed by matter domination.
Very sketchily, simulations of the evolution of a network of strings show that 
no matter what is the initial density of strings, within few Hubble times 
they reach a scaling solution characterised by few long open strings per Hubble
volume whose length stretches with the horizon, and a large number of small 
closed strings \cite{Vilenkin:book}. A network of long open strings has a 
negative effective pressure $p_{op}=-\rho_{op}/3$, thus it redshifts more 
slowly than both radiation and non-relativistic matter, but it does not 
overclose the Universe as open strings loose energy by chopping off small loops
which eventually decay by emission of gravitational radiation.
At an effective level one can write down a continuity equation which takes
into account both the Universe expansion and the energy lost into loops
\be \label{sceq}
\dot\rho_{op}=-3H(\rho_{op}+p_{op})-\lambda\frac{\rho_{op}}{L}=-2H\rho_{op}-
\lambda\frac{\rho_{op}}{L}\,,
\ee
where $0<\lambda<1$ is a number taking account of the possibility of a 
non-unity value of the probability of reconnection. For a generic Hubble 
parameter $H=\beta/t$ and $\rho\propto\mu/L^2$ eq.~(\ref{sceq}) admits the 
\emph{scaling} solution
\be
L=\frac\lambda{2(1-\beta)}t\,,
\ee
allowing the string network to \emph{track} the dominant energy source of 
the Universe and to make up a constant fraction of it.

String loops have typical size $l_{cl}$ and density $n_{cl}$ roughly given by 
\be \label{lenden}
l_{cl}\simeq\alpha t\,,\qquad n_{cl}\simeq \alpha^{-1}t^{-3}\,.
\ee
The value of $\alpha$ is not known, as an indication we can take the value 
given in \cite{Bennett:1987vf} $\alpha\simeq \Gamma G_N\mu$ where 
$\Gamma\simeq 50$, so that it can well be that $\alpha\ll 1$ (we remind that 
smoothness of cosmic microwave background requires $G_N\mu<10^{-7}$).
A relation similar to (\ref{lenden}) between $l_{op},n_{op}$ and 
$t$ can be expected to hold for long, open strings with $\alpha\simeq 1$.
Given (\ref{lenden}) the number of reconnection events between open strings
per unit of spacetime volume can be estimated to be
\be \label{denenc}
\nu_{op-op}(t)=n^2_{op} v \sigma_{op-op}\simeq Pt^{-4}\,,
\ee
where the cross section $\sigma_{op-op}$ has been estimated as 
$\sigma_{op-op}\simeq Pl_{op}^2$, with $P$ the reconnection probability, see 
(\ref{velprob}), and relativistic velocities has been considered.
Had we considered instead the interaction between open and closed strings
we would have obtained
\be
\nu_{op-cl}(t)=n_{op}n_{cl} v \sigma_{op-cl}\simeq P t^{-4}\alpha^{-1/3}\,,
\ee
which is slightly higher than (\ref{denenc}), where it has been used that the
average distance traveled by a string before meeting another one is 
$\sim n^{-1/3}_{cl}$.

We want now to relate the time variable $t$ to the redshift factor $(1+z)$ as
in \cite{Damour:2001bk}.
Let us consider the usual FRW metric
\be
ds^2=-dt^2+a^2(t)\pa{dr^2+r^2d\Omega^2}
\ee
the comoving distance $r$ between an object emitting light at a time $t(z)$,
where $(1+z)$ denotes the redshift factor, and the observer at present time 
$t_0$ is given by
\be
r(z)=\int_{t(z)}^{t_0} \frac{dt'}{a(t')}\,.
\ee
Parameterising the scale factor as 
\be
a=\left\{\ba{cl}a_{eq}\pa{t/t_{eq}}^{1/2} & t_i<t<t_{eq}\,,\\ 
a_0\pa{t/t_{0}}^{2/3}& t_{eq}<t<t_0\,,\ea\right.
\ee
where $t_i$ mark the epoch of the beginning of radiation domination and
$t_{eq}$ denotes the time of matter-radiation equality, the redshift-comoving 
distance relation is obtained
\be
r(z)=\frac{3t_0}{a_0}\times\left\{\ba{lc} 
\ds\pa{1+z_{eq}}^{-1/2}\paq{
\pa{1+z_{eq}}^{1/2}-1+\frac 23\frac {z-z_{eq}}{1+z}} & z_i>z>z_{eq}\\ 
\ds\frac{\pa{1+z}^{1/2}-1}{\pa{1+z}^{1/2}}& z_{eq}>z>0\ea\right.\,,
\ee
which, neglecting numerical factors, can be well interpolated by 
\be \label{distz}
r(z)\simeq\frac{t_0}{a_0}\frac z{1+z}\,.
\ee
The previous formulae allow us to write the volume element as
\be \label{volume}
dV(z)=4\pi a^3(t_0)r^2dr\simeq 10^2 t_0^3\frac{z^2}{(1+z)^{13/2}}
\frac 1{\pa{1+z/z_{eq}}^{1/2}}dz\,.
\ee
Let us introduce $N(z)$, the reconnection rate for events occurring at a
given redshift $z$. Its differential is given by
\be \label{rate}
dN=(1+z)^{-1}\nu(z)z\frac{dV}{dz}d\ln z\simeq 
\frac{10^2}{t_0}z^3\frac{\pa{1+z/z_{eq}}^{3/2}}{\pa{1+z}^{3/2}}
d\ln z\,,
\ee
where the time-redshift relation 
\be \label{tdiz}
t(z)=t_0(1+z)^{-3/2}(1+z/z_{eq})^{-1/2}
\ee
has been used.
The observer at $r=0$ sees a burst with duration $\Delta t_0=\Delta t\,(1+z)$, 
being $\Delta t$ the proper time burst duration. By 
comparing $(\Delta t_0)^{-1}$ with $dN/d\ln z$ we can verify if the signals 
actually overlap to make a continuous background, or if they are resolved in 
time. Since $N(z)$ is a growing function of $z$, it turns out that for $z_*>z$ 
the signal is made by a collection of bursts whereas signals originated at
higher redshift than $z_*$ combine into a stochastic background. Comparing 
(\ref{rate}) with $\Delta t_0$ one obtains
\be \label{zstar}
z_*\simeq 10^{10}\pa{\tan\gamma}^{1/8}\alpha^{1/4}
\pa{\frac{\al^{1/2}}{\rm{TeV}^{-1}}}^{-1/4}.
\ee
The average over $\gamma$ introduces a numerical factor $\simeq 0.9$.
In the case of continuous background, i.e. for events generated at redshift 
$z>z_*$, taking into account the cosmological redshift, the spectral density 
function can be estimated as
\be \label{sh}
S_h(f)=\int^{z_{in}}_{z_*}|\tilde h(f(1+z'))|^2 
\frac{dN(z')}{d\ln z'}d\ln z'\,,
\ee
where $z_{in}$ stands for the red-shift of the set in of the scaling solution 
and $\tilde h(f(1+z))$ stands for a sum over metric polarisation and average 
over directions $\hat n$ of $\tilde h_{\mu\nu}(f(1+z),\hat n)$.
Since $dN(z)/d\ln z\propto z^3$ at large $z$, the integral in eq.(\ref{sh})
can be approximated by the value of the integrand at the upper limit of 
integration region
\be
S_h(f)=|\tilde h(f(1+z_{in}))|^2\frac{10^2}{t_0}\frac{z_{in}^3}{z_{eq}^{3/2}}
\,,
\ee
where $\tilde h(f)$ can be read from (\ref{spectralh}) with the aid of 
(\ref{shdt}) after averaging over orientations. With respect to the signal
from an astrophysical source given by 
(\ref{spectralh}) there is a factor $10^2z_{in}^3/(t_0z_{eq}^{3/2})$ instead
of $\Delta t_0\simeq q^{-1/2}(1+z)$, leading to an enhancement for high enough 
redshift, but again the final result is negligible
\be \label{estim}
S_h(f)\simeq 10^{-144}{\rm Hz}^{-1}e^{-\pi^2f^2\al z^2_{in}}
\pa{\frac{G_N\mu}{10^{-7}}}^2\pa{\frac{\al}{\rm TeV^{-2}}}^2
\pa{\frac{z_{in}}{10^{10}}}^3\,,
\ee
where an average over D-string meeting angle $\gamma$ has been performed
and the distance $x$ from the source has been expressed as a function of 
the red-shift according to (\ref{distz}).
Moving to higher redshift brings competing effects: a higher distance from
the source, see eq.(\ref{distz}) and a decreased integration volume, 
see eq.(\ref{volume}), both of which tend to suppress the signal, and
increased event rate, see eq.(\ref{denenc}), which tends to increase the 
signal and it is actually the dominant effect.
We remind the relation between redshift $z$ and temperature $T$ during 
standard cosmological evolution
\be \label{Tzconv}
z\simeq 10^9\pa{\frac{T}{\rm MeV}}\,.
\ee

Actually before the scaling solution sets in, the density can be higher than 
(\ref{lenden}) and consequently the rate of string encounter higher 
than (\ref{denenc}). One could consider the very extreme case of an initial
density of cosmic D-strings 
\be \label{nex}
n_{ex}\sim \mu^{-3/2}{\al}^{-3}
\ee 
at the epoch of cosmic string creation (inflation ending), characterised by a 
redshift $z_{ex}$. Such a density would correspond to a Hagedorn energy 
density $\rho_h\sim\al^{-2}$, which is generally a limiting energy density for 
open string systems. The corresponding rate of reconnection events per volume 
is
\be \label{nuex}
\nu^{(ex)}_{op-op}\sim \mu^{-2}\al^{-4},
\ee
while the standard scaling solution is recovered in a few Hubble times. 
The highly non standard form of the D-brane tachyon action, which suppresses 
gradients of the field, allows us to assume a far higher density of defects
than the standard Kibble argument would allow as we will see in the next 
section, see \cite{Barnaby:2004dz}. 
With this assumption we can reconsider the integral (\ref{sh}), noting that it
will be dominated by the early epoch $z'\sim z_{ex}$ thus giving, analogously 
to (\ref{estim}),
\be
S_h(f)\simeq 10^3\frac{(G_N\mu)^2 \al^2}{x^2(z_{ex})}e^{-\pi^2f^2\al z_{ex}^2}
\frac {10^2t_0^3z^{-5}_{ex}z_{eq}^{1/2}}{\mu^2\al^4}\simeq
10^{-91}{\rm Hz}^{-1}e^{-\pi^2f^2\al z_{ex}^2}
\!\pa{\frac{\al}{\rm TeV^{-2}}}^{-2}\!\!\!\pa{\frac{z_{ex}}{10^{10}}}^{-5}
\!\!\!\!\!\!.
\ee
which is still way too small for detection.
It has also to be noted that the Kibble density is smaller than the extremal 
density $n_{ex}$ given in (\ref{nex}) only if
\be \label{enhan}
\mu<G_N^{-1/3}l_s^{-4/3}\,.
\ee

With the process considered so far one cannot obtain a large enough result, as
an inherently microscopic mechanism is involved, which affects a portion of the
string of size the order of the string length in a huge Universe.
For the gravitational emission to give a sizeable effect we need it to be 
produced in a coherent motion of all its world volume.
In sec.\ref{annhi} we move to the study of systems of D-string
$\rm{\overline D}$-string which annihilate, but before doing that we
complete the analysis of recombination by describing two almost antiparallel 
D-strings, which is a system close to the D $\rm \overline D$-string case.

\subsection{Almost antiparallel case}

\subsubsection{System dynamics}

If the system of two D-strings is close to a brane-anti brane pair it can
still be analysed through a Yang Mills low energy effective action, but a 
scalar tachyon has to be added to the spectrum, as it comes from the lowest 
lying excitations of the string stretched between the brane and the anti-brane
\cite{Hashimoto:2003xz}.The action is then
\be
\az=-\mu (2\pi\al)^2\int dt dx\paq{{\rm Tr}\pa{\frac 14 F_{\mu\nu}F^{\mu\nu}+
\frac 12D_\mu\phi^iD^\mu\Phi^i+\frac 12 D_\mu TD^\mu T-\frac{m^2}2T^2}}\,,
\ee
with $m^2=-1/(2\al)$\,.
Now the expansion has to be taken around the background (again 
$q\equiv\tan\gamma/(\pi\al)$)
\be \label{backbab}
\Phi_i^{(b)}=\delta_{i,2}qx\,\si^3\,,\qquad A_\mu^{(b)}=0\,,\qquad T^{(b)}=0\,,
\ee
and the fluctuations
\be \label{pertbab}
\Phi_i^{(f)}=\delta_{i,2}\xi(t,x)\,\si^3\,,\qquad A_\mu^{(f)}=0\,,\qquad
T^{(f)}=\tau(t,x)\si^2\,.
\ee
In terms of the ans\"atze (\ref{backbab}) and (\ref{pertbab}) the Lagrangian,
up to lowest order interaction, can be written as
\be
\lag\propto \frac 12 \dot\tau^2-\frac 12{\tau'}^2-\frac 12q^2-
\frac 12\pa{q^2x^2+2qx\xi}\tau^2+\frac 1{4\al}\tau^2+\frac 12 \dot\xi^2-
\frac 12{\xi'}^2\,.
\ee
The analysis follows straightforwardly as in the previous case to obtain
the equations of motion
\be
\ba{rcl}
\ddot \tau-\tau''+q^2x^2\tau-\tau/(2\al)&=&0\,,\\
\ddot\xi-\xi''+\tau^2qx&=&0\,.
\ea
\ee
Again the unstable mode solution is
\be
\ba{c}
\tau(t,x)=C_0e^{\beta t}e^{-qx^2/2}\,,\\ \ds
\xi(t,x)=C_0^2e^{2\beta t}\frac e4\frac{\sqrt\pi}\beta\paq{\cosh(2\beta x)
-\frac 12\pa{e^{2\beta x}\erf(1+\beta x)+e^{-2\beta x}\erf(-1+\beta x)}}\,,
\ea
\ee
where $\beta\equiv \sqrt{1/(2\al)-q}$.
\begin{figure}
  \centering
  \includegraphics[width=.8\linewidth]{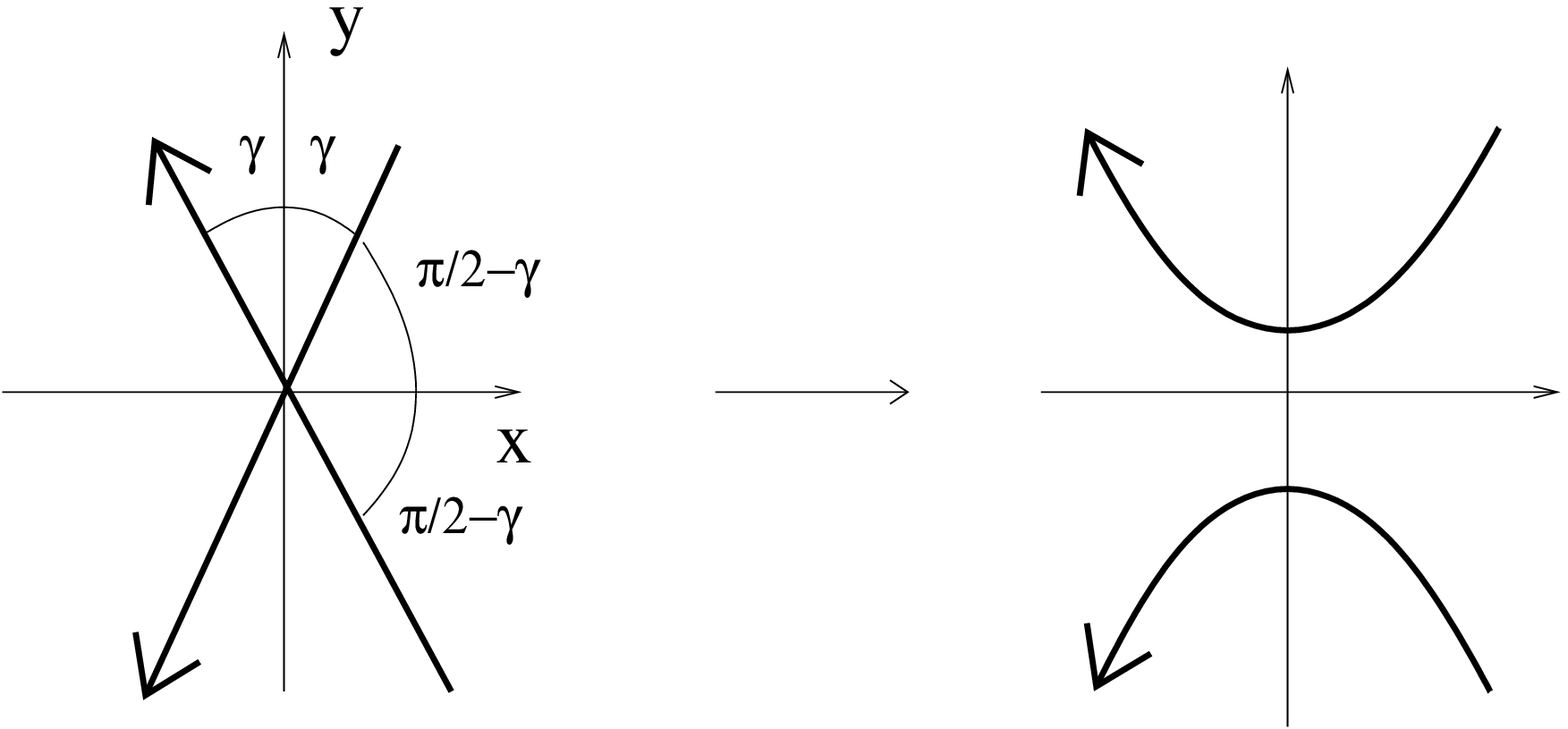}
  \caption{Two D-branes at an angle $\pi-\gamma$ before and after 
    recombination.}
  \label{dantidga}
\end{figure}

\subsubsection{Gravitational radiation}

From the dynamics exposed in the previous section the energy momentum tensor 
can be estimated to be
\be
\ba{cl}
T_{\mu\nu}=&\ds 2\mu\pi^2\al^2\pa{
\ba{ccc}
-q^2/4-\beta^2\xi'/2 & -\beta^2\dot \xi/2\\
-\beta^2\dot\xi/2 &\tau^2\beta^2/2 -\beta^2\xi'/2 -q^2/4
\ea} \\
\ds &\ds\times\paq{\delta\pa{y-qx\pi\al}+\delta\pa{y+qx\pi\al}}\,.
\ea
\ee
It is then straightforward in full analogy with sec.\ref{emt1} to estimate
the gravitational radiation, obtaining the same result.

\section{String-anti string annihilation}

\label{annhi}
We have seen that the recombination process of D-strings described in 
sec.~\ref{recmb} is not a phenomenologically interesting source of 
gravitational radiation. This was due to the smallness of the size of the 
interaction region, which was of the order of the string length. In general 
interesting sources of gravitational radiation are related to coherent motion 
of large and extended portions of mass. 

We thus now turn to study a process in which two 
D-strings meet and \emph{annihilate}, thus enabling a much larger reservoir of 
energy to be converted into gravitational radiation. Such systems have been 
studied in detail in a full string theory context for different configurations
of the tachyon describing the instability \cite{Sen:2004nf}.\\
In particular it has been shown that the tachyon can acquire different spatial 
profiles, triggering the decay of the original system into different final 
states. For instance a trivial tachyon profile in the space dimensions will 
lead the brane anti-brane system towards the so called \emph{tachyon matter} 
which will be discussed in sec.\ref{remnt}, whereas a kink profile will let
a D$p$ $\bar{\rm D}p$-brane system decay into a D$(p-1)$-brane, a vortex 
configuration into a D$(p-2)$-brane and so on.
We show next that in the case of a trivial spatial configuration, important
conclusions can be drawn for the production of gravitational radiation.

\subsection{Recombination and annihilation}

For ordinary phase transitions the presence of defects is related to the
global configuration of some field triggering the transition. In an expanding 
Universe with Hubble rate $H$, the defect separation $d$ must be smaller than
the Hubble length $H^{-1}$, as any super-horizon correlation would violate 
causality, thus leading to the lower bound of one defect per Hubble volume 
(\emph{Hubble density}), and the actual number of defects per Hubble volume is 
as large as a few.

However the density of defects can be much larger than the Hubble one as this 
picture does not apply to the tachyon condensation, whose dynamics is
described by the Lagrangian 
\be \label{vdit}
\az=-\mu\int d^{p+1}x V(T)\sqrt{1+\dpa_\mu T\dpa_\nu T g^{\mu\nu}}\,,
\ee
where the $V(T)$ is a symmetric, positive potential, which vanishes
asymptotically for $T\to \pm\infty$ as $e^{-T}$ \cite{Sen:2002innu} 
(it is not very important here the exact form of such potential).

As shown in \cite{Barnaby:2004dz}, the non-standard form of the kinetic terms 
implies that once the tachyon has rolled to large values, gradients are
exponentially suppressed, essentially eliminating the restoring force which
would tend to eliminate gradients within a casual volume, thus allowing a
higher defect density than the Kibble value.\\
We assume that the density $n_{ex}$ as in (\ref{nex}) can be
achieved, corresponding to an energy density of the order of the Hagedorn 
scale, which is a limiting energy density for open strings \cite{Atick:1988si}.

Actually in the evolution of an ordinary string network, the initial
conditions are not crucial as the network reaches the scaling solution in a
few Hubble time, long strings loose energy by chopping off small loops, which 
in standard simulations then looses energy by emitting gravitational wave
until disappearance. If we take the extremal case of Hagedorn energy density 
like in the end of sec.~(\ref{reccr}) the long string inter-separation is 
$d_{ex}\sim \mu^{1/2}\al$ rather than the Kibble value 
$d_K\sim H^{-1}\simeq G_N^{-1/2}\mu^{-1}$, which is eventually reached after 
few Hubble times.

The high starting density of strings is anyway not enough in itself to produce 
a large amount of gravitational wave if the usual picture is assumed, i.e. that
small loop radiates as in \cite{Damour:2001bk} via the formation of cusps and
kinks. Again this is due to the smallness of the physical size of the radiating
objects, in this case the small loops, whose typical size is some orders of 
magnitude smaller than the Hubble scale, see eq.~(\ref{lenden}), where 
$H\sim \sqrt{G_N\rho}\simeq l_{Pl}/l_s^2$, taking for $\rho$ the extremal 
Hagedorn density. \footnote{Here we have a limiting 
density different from \cite{Barnaby:2004dz}, where the energy scale relative 
to the string tension and to the inverse distance are identified, as no 
Hagedorn limit}.
In this picture the emission is originated in the early Universe, when the 
size of cosmic string is small as they had no time to stretch with
cosmological expansion, so a small amount of radiation is produced.
In \cite{Damour:2001bk} for instance, a typical signal close to experimental 
sensitivity at a frequency $f$, with rate $\dot N$, originated at redshift $z$ 
is obtained from a kink in a cosmic loop of size $l_{cl}$ roughly given by
\be
l_{cl}\simeq 10^{20}{\rm m}\pa{1+z}^{-3/2}\pa{1+z/z_{eq}}^{-1/2}
\pa{\frac{\alpha}{10^{-6}}}\,,
\ee 
and the rate $\dot N$ of such an event at a given frequency $f$ is roughly
given by
\be
\dot N=2\times 10^{-6}{\rm year}^{-1}\
\frac{z^3\pa{1+z/z_{eq}}^{11/6}}{(1+z)^{7/6}}\pa{\frac\alpha{10^{-6}}}^{-8/3}
\pa{\frac f{\rm kHz}}^{-2/3}\,.
\ee

\subsection{Annihilation with spatially trivial tachyon profile}

\label{remnt}
So far we have dealt with an effective field theory description of D-string 
$\overline{\rm D}$-string annihilation, the excitations of the branes have been
described as classical fields subject to ordinary second order differential 
equations.
  
On the other hand the annihilation process admits an \emph{exact} description 
in terms of the fundamental string theory, i.e. in terms of the
two-dimensional world-sheet theory, by adding to the standard world sheet 
action
\be
\az_{ws}=-\frac 1{4\pi}\int d\tau d\sigma\pa{\frac 1\al\dpa_a X^\mu\dpa^aX^\nu
g_{\mu\nu}+\bar\psi^\mu\rho^\alpha\dpa_\alpha\psi^\nu g_{\mu\nu}}\,,
\ee
a boundary term
\be \label{boundef}
\az_b\propto \lambda\int d\tau \psi^0e^{X^0/(2l_s)}\otimes\si_1\,.
\ee
The world-sheet (disk) partition function $Z_{ws}$, with the world sheet 
couplings interpreted as space-time fields, corresponds to the space time 
action $\az$ \cite{Fradkin:1985ysCallan:1985ia}, given by 
\be \label{part}
\az=Z_{ws}\propto \int d^px\sqrt{-g}\int\paq{dX^\mu}\paq{d\psi^\mu}
e^{-\az_{ws}-\az_b}\,.
\ee

The boundary term describes the condensation of the tachyon to a profile of
the type 
\be
T=\lambda e^{t/(\sqrt 2 l_s)}\otimes\sigma_1\,.
\ee
Giving an expectation value to the tachyon shifts the mass squared of the open 
string modes by an amount $\Delta m^2\propto\langle T^2\rangle$, eventually 
dropping them out of the physical spectrum as the tachyon increasing 
expectation value makes them more and more massive.

A D-brane in the free theory can be described by a specific boundary state
$\ket{B}$, the $\ket{B}\to\ket{B}$ vacuum amplitude is written as the  
so called \emph{cylinder} amplitude
\be
A_{BB}=\bra{B}P\ket{B}\,,
\ee
where $P$ is the closed string propagator. When the internal particle running
in the propagator goes on shell the amplitude $A_{BB}$ picks an imaginary
part.
Once written the expansion of $\ket{B}$ into a set of complete states as 
\be
\ket{B}=\sum_fU(\omega_f)\ket{f}\,,
\ee
by the optical theorem we have
\be \label{opt}
{\rm Im} A_{BB}=\sum_f \frac 1{2\omega_f}|U(\omega_f)|^2\,,
\ee
where $\omega_f$ is the energy of the state $f$.
In the unperturbed theory, where the tachyon has vanishing expectation value, 
the brane boundary state can be written as 
\be
\ket{B}=\mathcal{O}\ket{\Omega,k}\,,
\ee
where $\mathcal{O}$ is an operator build out of fermionic and bosonic matter 
oscillators and ghost oscillators and $\ket{\Omega,k}$ is the Fock vacuum with 
momentum $k$. The specific form of the vacuum will not be needed here, as we
shall simply quote known results, see \cite{Sen:2002innu,Larsen:2002wc}.
Taking the initial condition to be that of D-strings, the addition of the 
boundary term (\ref{boundef}) turns $\ket{B}$ into 
$\ket{B}_T$, whose energy momentum tensor is obtained from (\ref{part}) 
via (\ref{emtdef})
\be
T_{\mu\nu}=2\mu\ {\rm diag}\pa{-1,g(t),0,\ldots,0}\,,
\ee
with
\be
g(t)=\frac 1{1+2\pi^2\lambda^2e^{\sqrt 2t/l_s}}\,.
\ee
The boundary state $\ket{B}_T$ at $t\to -\infty$ corresponds to $\ket{B}$, and 
at $t\to\infty$ tends to a state without the original D-strings but 
endowed with the same amount of energy and vanishing pressure. 

Applying the optical theorem as in (\ref{opt}) one can find, see 
\cite{Shelton:2004ij,Karczmarek:2003xm},
that the expansion coefficient $U_T(\omega_f)$ relative to $\ket{B}_T$ turns 
out to be just the Fourier transform of $g(t)$, which is \footnote{This is 
actually the coefficient for the expansion over Neveu-Schwarz states, which is 
the interesting sector for gravitons.}
\be \label{udiom}
|U_T(\omega_f)|=\frac{\pi/\sqrt 2}{\sinh(\pi l_s\omega_f/\sqrt 2)}\,.
\ee
The imaginary part of ${}_T\langle{B}\ket{B}_T$ is related to the average 
numerical density $\bar n$ and energy density $\bar \rho$ of particles emitted 
by
\be
\bar n&=&\sum_f\frac 1{2\omega_f}|U(\omega_f)|^2\,,\\
\bar\rho&=&\sum_f\frac 12|U(\omega_f)|^2\,,
\ee
where the sums run over a basis of closed string states, implicitly including 
both the sum over excitation level $N$ of the closed string and the momenta of 
the spatial directions transverse to the brane.
For large $N$ the degeneracy of states $\mg(N)$ of a closed string in $D$ 
dimensions is 
\be \label{gdin}
\mg(N)=N^{-(D+1)/2}e^{4\pi\sqrt{\frac{D-2}6}\sqrt N}\,,
\ee
and here we assume $D=10$ from superstring theory.
Given the degeneracy of the states and the mass formula for closed strings
\be \label{massfor}
\al m^2=4\pa{N-\frac 12}\,,
\ee
the exponential in eq.~(\ref{udiom}) and (\ref{gdin}) combine
to leave a power law spectrum for the radiated energy according to 
\be \label{en}
\bar\rho&\propto&\int dE\ E^{-p/2}\,, 
\ee
for D$p$-branes, with a total emitted energy which is ultraviolet divergent for
$p\leq 2$.
Of course the emitted energy cannot be larger then the mass of the initial
brane, so back-reaction is expected to set in and cut off the emission, with
the result that the total radiated energy is finite and of the order of
the initial mass of the brane. In particular cutting off the integral in 
eq.~(\ref{en}) at $E_{cut}\sim\mu l_{op}$ would give all the energy into closed
string modes, emitting most of their energy into very massive strings.

Actually eq.(\ref{en}) seems to suggest that for $p>2$ the amount of energy of
the initial D$p$-brane going into closed string modes is rather small. However
the decay described by the boundary term (\ref{boundef}) is spatially 
homogeneous and it is not expected to realistically describe the decay
process. Perturbations with a non trivial spatial profile can still be 
tachyonic and thus contribute the decay. We should expect the decay to take 
place rather inhomogeneously and the original D$p$-brane can be thought of
decomposing into a collection of small patches. In each of this small patches,
disconnected one from the other, the behaviour of a D$p$-brane is similar to 
that of a collection of D0-branes \cite{Sen:2004zm}.

\subsection{Gravitational mode production}

The problem of gravitational emission has now been traded from the study of
annihilating D-branes to the one of decaying of a highly massive (hence highly 
excited) closed fundamental string.
Two-body decays of F-strings have been studied in literature for both the 
bosonic string \cite{Manes:2001cs} and the superstring 
\cite{Chen:2005ra,Iengo:2006gmif}.
The decay of a closed string into $n$ closed strings is a perturbative process
whose amplitude $\Gamma\propto g_c^n$, thus in the perturbative regime 
$g_c\ll 1$ the dominant process is two-body decay. 

Let us consider the amplitude decay $A$ of a \emph{specific} initial closed 
string state $\phi_N$ of mass $M\simeq 2\sqrt N/l_s$ into two \emph{given} 
final states $\phi_{N_1},\phi_{N_2}$ of masses respectively 
$m_{1,2}\simeq 2 \sqrt{N_{1,2}}/l_s$.
The inclusive decay width of a generic fundamental string excited at mass level
$N$, into two \emph{generic} final states at mass level $N_1,N_2$ is obtained 
by integrating the amplitude squared over the available final states, and  
averaging over initial states at mass level $N$.  We assume to have 
$d=3+1$ large ordinary dimensions and $d_c$ small ones, and a highly energetic 
(larger than the string scale) closed fundamental string which can both wrap 
and have momentum around the small dimensions. Phase space integration
includes a sum over winding and momenta. 

The differential decay width $d\Gamma_{cl\to m_{1,2}}$ is then given by 
averaging over initial states and summing over final ones 
\cite{Manes:2001cs,Chen:2005ra}
\be \label{dgamma}
d\Gamma_{cl\to m_{1,2}}=\frac 1{2M\mg(N)}\sum_i|A|^2d\Phi_{f_1,f_2}\simeq
\frac{g_c^2}{2M}\pa{N-N_1-N_2}^2\frac{\mg(N_1)\mg(N_2)}{\mg(N)}
d\Phi_{f_1,f_2}\,.
\ee
The two-body phase space in $d=3+1$ dimensions is 
\be
d\Phi^{(3+1)}_{f_1,f_2}=\frac{k}{16\pi^2M}d\Omega
\ee
where $k$ is the modulus of the momentum of the decay states and $\Omega$ the 
solid angle. Considering for simplicity flat internal compact dimensions, the 
inclusion of momentum and winding modes in the mass formula (\ref{massfor}) 
adds a sum over momenta and windings in the compact directions. The
closed string mass formula which generalises (\ref{massfor}) in the presence of
discrete momenta and windings is
\be
\al m^2=4\pa{N-\frac 12}+\sum_i\pa{\al\frac{n_i^2}{R_i^2}+
w^2_i\frac{R_i^2}\al}\,,
\ee
which for large energies can be written as
\be
l_sm\simeq 2\sqrt N+
\frac 12\sum_i\pa{\frac{n_i^2l_s}{R_i^2m}+\frac{w_i^2R_i^2}{l_s^3m}}\,.
\ee
This allows to approximate the summations over $n_i$ and $w_i$ as Gaussian 
integrals in the case respectively $m\gg l_sR_i^{-2}$ and $m\gg R^2_i/l_s^3$.
Using the asymptotic formula for the density of states (\ref{gdin})
the amplitude (\ref{dgamma}) can be integrated to \footnote{We assume that 
the closed strings have enough energy to possess momenta and windings along 
the compact dimensions, if this were not the case the right hand side of 
(\ref{gamma}) should be divided by $3^{1/4}\sqrt{l_sm}\frac{R_i}{2l_s}$
for each direction $i$ along which the momentum modes are not excited and
$3^{1/4}\sqrt{l_sm}\frac{l_s}{2R_i}$ for each direction with 
unexcited winding modes.}
\be \label{gamma}
\Gamma_{cl\to m_{1,2}}\simeq
\frac{2\pa{2\pi/a}^{d_c}}{\pi M^{2-d_c}l_s^{3+D-d-2d_c}}e^{-al_sE}
\pa{\frac{m_1m_2}{M}}^{-(D-d_c+1)}(m_1m_2)^2
\pa{\frac{2Em_1m_2}M}^{\frac{d-3}2}\,,
\ee
where $a=4\pi/\sqrt 3$, $d_c$ the number of compact dimensions along which
the closed strings have momenta and windings, the kinetic energy of the decay
objects
\be
E\equiv M-m_1-m_2\simeq \frac{M}{2m_1m_2}k^2
\ee
has been introduced and the approximate relation
\be
N-N_1-N_2\simeq\al M\sqrt{k^2+m_1^2}-m_1^2\simeq\al m_1m_2
\ee
has been used.

Let us call $\Gamma_m$ the total rate of production of a string of mass 
$m$,  which is obtained by summing $\Gamma$ over all values $N_2$. The sum 
over $N_2$ can then be trade for a continuous integral over $E$ with the 
substitution $\sum_{N_2}\to -\al m_2dE$, obtaining
\be
\Gamma^{(m)}_{cl\to m_{1,2}}\propto g_c^2
\frac{\pa{l_sM}^{D-d+1}}{l_s^2m}m_R^{-(d-1)/2}\int E^{(d-3)/2}e^{-al_sE}dE\,,
\ee
where $m_R\equiv m(M-m)/M$. Since in an interval $\Delta N$ corresponds to 
$\al m\Delta m/2$, the density of states per mass level is $\rho(m)=\al m/2$, 
thus we can define the differential decay width per mass interval
\footnote{Note taht we obtain a dependance on $M$ and $m$ different from 
\cite{Chen:2005ra}. This is basically because here we have summed over 
Kaluza-Klein momenta and winding modes of the decaying objects, thus obtaining
the expected dependence of (\ref{dgammam}) on $m$ insteading of the weird
$m^{-(D-1+d_c)/2}$ of \cite{Chen:2005ra}.}
\be \label{dgammam}
\frac{d\Gamma_{cl\to m_{1,2}}}{dm}=\Gamma^{(m)}_{cl\to m_{1,2}}\rho(m)\propto 
g_c^2M\pa{l_sM}^{D-d}\pa{l_sm_R}^{-(d-1)/2}\,.
\ee
The differential decay rate (\ref{dgammam}) clearly shows that light strings 
are more easily produced than heavy ones.
Thus we can safely assume that a consistent fraction of the initial rest mass 
is converted into massless closed strings, i.e. radiation.

We are then lead to consider the amplitude inclusive over $m_2$ and with 
$m_1=0$ which is \cite{Manes:2001cs}
\be \label{decayto0}
d\Gamma_{cl\to rad}\simeq g_c^2
\frac{e^{-al_sk}}{1-e^{-al_sk}}l_sM(l_sk)^{d-1}dk\,,
\ee
similar to a black body spectrum, leading to a total decay rate 
\be
\Gamma_{cl\to rad}\simeq g^2_c M\,.
\ee
Thus in a time much shorter than the string scale a D-string and a 
$\overline{\rm D}$-string decay efficiently into radiation.\\
If it is so one of the two decay objects can still be massless from the 
ten-dimensional point of view but it will be massive from the effective 
4-dimensional theory.
In this case the decay width (\ref{decayto0}) should have been modified by the
inclusion of an additional integral 
\be \label{spekk}
d\Gamma_{cl\to KK}= g_c^2
\frac{e^{-al_s\sqrt{k^2+m^2}}}{1-e^{-al_s\sqrt{k^2+m^2}}} 
l_sM(l_sk)^{d-1}dk\prod_i^{d_c} R_idk_i\simeq g_c^2 e^{-al_s\sqrt{k^2+m^2}}
m^{d_c-1}dm\,.
\ee

Then integration over the momenta in the compact dimensions have the effect of 
multiplying the differential decay width by a factor $\prod R_i/l_s$, for
$R_i\gg l_s$, because of the opening up of new decay channels, and it does not
modify the result (\ref{decayto0}) for $R_i\sim l_s$.
  
\begin{figure}
  \centering
  \includegraphics[width=.8\linewidth]{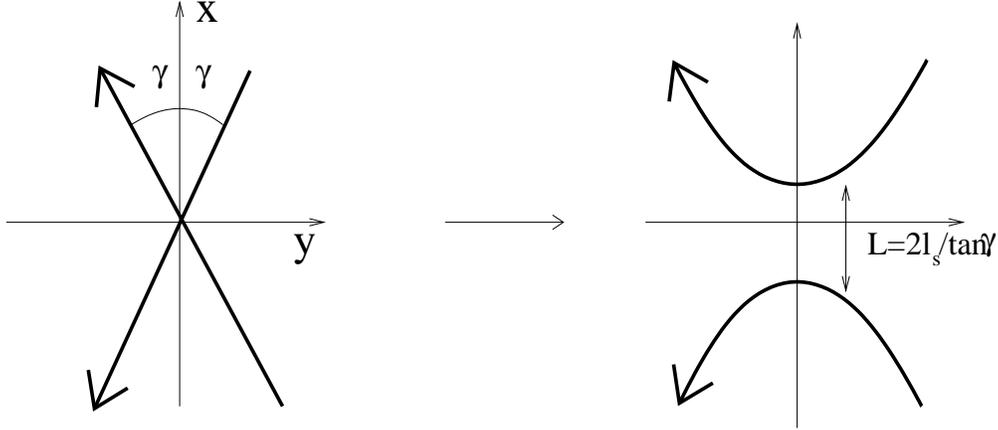}
  \caption{Recombination of two almost antiparallel D-strings. It is 
    highlighted the region of space where tachyon condensation occurs.}
  \label{decay}
\end{figure}

If two D-strings are not exactly anti-parallel but meet at an angle $2\gamma$, 
as in fig.~\ref{dantidga}, the tachyon condensation makes the D-string 
disappear not over the entire world-volume of the D-strings, but only as long 
as the D-strings are separated by a distance shorter than $2l_s$, which for 
strings meeting at angle $2\gamma$ happens for a region of size 
$L_x=2l_s/\tan\gamma$, see fig.~\ref{decay}.

The energy density in gravitational waves $\rho_{gw}$ is related to the 
spectral density function by \cite{Maggiore:1999vm}
\be \label{rhosh}
G_N\rho_{gw}=\frac\pi 2\int_0^\infty df f^2 S_h(f)\,.
\ee
The energy a single annihilation of D-branes make available for radiation
can be estimated as $\mu L_x$ which eventually expands spherically. 
From (\ref{decayto0}) and (\ref{rhosh}) one can then compute the strain
due to a gravitational wave emitted at a distance $x$ from us (here we neglect
red-shift for the moment, thus having $fl_s\ll 1$ at interesting frequencies)
\be
S_h(f) \simeq &\ds\frac 2\pi \frac{G_N\mu L_x\al}{x^2}
\frac{l_sf\ e^{-al_sf}}{1-e^{-al_sf}}\simeq 
10^{-108}{\rm Hz}^{-1}\frac{e^{-al_sf}}{\tan\gamma}
\pa{\frac{G_N\mu}{10^{-7}}}\pa{\frac \al{\rm TeV^{-2}}}^{3/2}\!\!\!
\pa{\frac{x}{\rm Mpc}}^{-2}\,.
\ee
Of course $S_h(f)$ cannot diverge for $\gamma\to 0$, as the naive estimation 
for $L_x$ does, but for small $\gamma$ one has to substitute $2l_s/\tan\gamma$
with $l_{op}$, the (open) cosmic string length. In this optimal case of
perfect anti-alignment one has 
\be
\ba{rl}
S_h^{(op)}(f)\simeq & \ds \frac 2\pi\frac{G_N\mu l_{op}\al}{x^2}
\frac{l_sf(1+z)e^{-al_sf(1+z)}}{1-e^{-al_sf(1+z)}}\simeq \\
&\ds 10^{-70}{\rm Hz}^{-1}e^{-al_sf(1+z)}
\frac{\pa{1+z}^{1/2}}{z^2\pa{1+z/z_{eq}}^{1/2}}\pa{\frac{G_N\mu}{10^{-7}}}
\pa{\frac \al{\rm TeV^{-2}}}\,,
\ea
\ee
where it has been used that $l_sf(1+z)\ll 1$, as a frequency $l_s^{-1}$ at 
an epoch characterised by a temperature $T=l_s^{-1}$, once red-shifted 
corresponds to $100$GHz today, assuming standard adiabatic evolution in 
between.

Assuming that the radiation is produced by an annihilation event during 
cosmological evolution when the scaling solution has set in, the expected rate 
of such an event is
\be
N=\frac Pt\simeq \frac P{t_0}(1+z)^{3/2}\pa{1+z/z_{eq}}^{1/2}\,.
\ee

It is clear that here again we have to assume an extremal initial density of
strings well above the Kibble density to hope to have a detectable signal.
Let us then assume an initial density of annihilation (\ref{nuex}) 
at an epoch characterised by a red-shift $z_{ex}$. Annihilation will thus take
place in a region of volume $\Delta V(z_{ex})$, see eq.(\ref{volume}), and for 
a time duration 
\bdm
\Delta t(z_{ex})=\left|\frac{dt}{dz}\right|_{z=z_{ex}}\Delta z\simeq 
\left|\frac{dt}{dz}\right|_{z=z_{ex}} z_{ex}
\edm 
to give
\be \label{shex}
\ba{rl}
\ds S_h^{ex}(f)=&
\ds \frac {2}{\pi f^2} G_N\frac{d\rho_{gw}}{df}\nu_{ex}\Delta V\Delta t=\\
&\ds\frac 2\pi\frac{G_N\mu l_{op}\al}{t_0^2z^2/(1+z)^2}\times
\frac{l_sf(1+z)e^{-al_sf(1+z)}}{1-e^{-al_sf(1+z)}}\times
\frac 1{\mu^2\al^4}\times\frac{t_0^4z^4}{(1+z)^9(1+z/z_{eq})}\simeq\\
&\ds 10^{50}\frac{z^2}{(1+z)^{17/2}(1+z/z_{eq})^{3/2}}{\rm Hz}^{-1}
\pa{\frac{\mu}{10^{-7}G_N^{-1}}}^{-1}\!\!\pa{\frac{\al}{\rm TeV^{-2}}}^{-3}
e^{-al_sf(1+z)}\,,
\ea
\ee
where we have considered the optimal case $L_x=l_{op}$.
Since the extremal case of a Hagedorn energy density has been assumed, 
$z_{ex}$ has to be high enough to accommodate such a high density phase.
Assuming standard adiabatic expansion after this extremal density phase the 
relation between the red-shift factor $z$ and the string scale is 
eq.(\ref{Tzconv})
\be \label{zexa}
z=10^{15}\pa{\frac{\al}{\rm TeV^{-2}}}^{-1/2}\,.
\ee
which takes the estimate (\ref{shex}) very far from both experimental 
sensitivity and phenomenological bound. 
In particular the nucleosynthesis bound requires that not too much radiation 
is present at the time of the synthesis of the primordial elements, in order 
not to spoil the beautiful agreement between theory and observation of 
primordial element abundances. Such bound constraints
\be
S_h(f)<10^{-51}{\rm Hz}^{-1}\pa{\frac{\rm{kHz}}f}^3\,.
\ee
Substituting (\ref{zexa}) into (\ref{shex}) we can obtain 
\be
S_h(f)=10^{-72}{\rm Hz}^{-1}\pa{\frac \mu{10^{-7}G_N^{-1}}}^{-1/2}
\pa{\frac\al{\rm TeV^{-2}}}^{-5/2}\pa{\frac {f}{10^{11} \rm Hz}}
\frac{e^{-al_sf(1+z)}}{1-e^{-al_sf(1+z)}}
\ee 
In our toy model the spectral strength can saturate the bound only around 
$10\div 100$GHz, i.e. at the end of the spectrum. This is hardly surprising as 
most of the energy of the F-string decay is concentrated for frequencies 
$f^{-1}\sim l_s(1+z_{ex})$.

On the other hand if Kaluza-Klein modes are created and if they are absolutely 
stable, they can rapidly overclose the Universe. In this case of extremal 
production their initial energy density is a non negligible fraction of the 
total energy density, so their decay is necessary otherwise they would start 
dominating as soon as they become non-relativistic. Their final abundance is
then set by the annihilation cross section \cite{Servant:2002aq}, which may
not be Planck scale, as gravity can be strong much below the Planck scale in
large extra dimension scenarios.\\
If instead we consider the standard scenario with the space-time rate of open 
string-open string encounters given by (\ref{denenc}), we can estimate the 
fraction of the critical energy density going into the production of massive 
gravity modes as follows. 
Let us suppose that each encounter give rise to the production of massive
modes for a total energy $E_{KK}=\mu L_x$, thus giving an energy density per 
Hubble volume $\rho_{KK}\simeq \mu L_x/t^3$. By normalizing it by the critical 
energy density $\rho_c=3H^2/(8\pi G_N)\simeq (8\pi G_N t^2)^{-1}$ and
integrating over the history of the Universe, using eq.(\ref{tdiz}), we have
\be \label{omkk}
\ba{rl}
\ds \Omega_{KK}(z_{nr})\equiv\frac{\rho_{KK}(z_{nr})}{\rho_c(z_{nr})}=&\ds
32\pi^2 G_N\mu\frac{l_s}{t_0}
\int_{z_{nr}}^{z_{in}}\pa{1+z'}^{1/2} \pa{1+z'/z_{eq}}^{1/2}dz\\
\simeq & \ds 10^{-50}\frac{z^2_{in}}{z_{eq}}\ds \pa{\frac{G_N\mu}{10^{-7}}}
\pa{\frac \al{\rm TeV^{-2}}}^{1/2}\,,
\ea
\ee
where $z_{in}$ denotes as in eq.(\ref{sh}) the time of the onset of the
scaling solution. Note that the integral in (\ref{omkk}) has been halted at 
$z_{nr}$, the red-shift at which the Kaluza-Klein modes become non 
relativistic. For $z_{nr}>z>z_{eq}$, $\rho_{KK}$ keeps earning a factor of $z$ 
over $\rho_c$ thus resulting into a present fractional energy density 
$\Omega_{KK}\equiv\Omega_{KK}(z=0)$ given by
\be
\Omega_{KK}=\Omega_{KK}(z_{nr})\frac{z_{nr}}{z_{eq}}\simeq 
10^{-11}\pa{\frac{G_N\mu}{10^{-7}}}\pa{\frac\al{\rm TeV^{-2}}}^{-1}.
\ee
where for simplicity we have assumed the mass of the Kaluza-Klein modes to be 
$\al^{-1/2}$, see eq.(\ref{spekk}), and substituted the relation (\ref{zexa})
for $z=z_{in}=z_{nr}$.\\
This result can be phenomenologically relevant for a still moderate value of
the string scale.

\section{Conclusions}

\label{concl}
Motivated by the recent revival in cosmic strings we have studied cosmic 
strings by modeling them as D-branes, the solitonic objects of string 
theory and computed the amount of gravitational waves emitted in recombination
processes which can take place in the moderately early Universe.
We have found that in general such processes are well below present 
sensitivities.
Still the process leading to string annihilation can result into a production
of gravity modes of fundamental strings which can be massive at an effective
4-D level and can thus have a cosmological effect. 
Such effect depends ultimately, in our simple model, on the string scale 
value, and it can be interesting even if the string scale is well below the 
Planck scale.
From a more theoretical point of view, this mechanism represents a new way for 
producing out of equilibrium, very massive particles.

\section*{Acknowledgments}

It is a pleasure to thank F. Dubath, R. Durrer, M. Kunz and M. Sakellariadou 
for very useful discussions. J.L.C. and E.P. whish to thank the theoretical 
physics department of the University of Geneva for kind hospitality during the
completion of part of this work.

\end{document}